\documentstyle[12pt]{article}

\catcode`@=11
\newcount\@tempcntc
\def\@citex[#1]#2{\if@filesw\immediate\write\@auxout{\string\citation{#2}}\fi
  \@tempcnta\z@\@tempcntb\m@ne\def\@citea{}\@cite{\@for\@citeb:=#2\do
    {\@ifundefined
       {b@\@citeb}{\@citeo\@tempcntb\m@ne\@citea\def\@citea{,}{\bf ?}\@warning
       {Citation `\@citeb' on page \thepage \space undefined}}%
    {\setbox\z@\hbox{\global\@tempcntc0\csname b@\@citeb\endcsname\relax}%
     \ifnum\@tempcntc=\z@ \@citeo\@tempcntb\m@ne
       \@citea\def\@citea{,}\hbox{\csname b@\@citeb\endcsname}%
     \else
      \advance\@tempcntb\@ne
      \ifnum\@tempcntb=\@tempcntc
      \else\advance\@tempcntb\m@ne\@citeo
      \@tempcnta\@tempcntc\@tempcntb\@tempcntc\fi\fi}}\@citeo}{#1}}
\def\@citeo{\ifnum\@tempcnta>\@tempcntb\else\@citea\def\@citea{,}%
  \ifnum\@tempcnta=\@tempcntb\the\@tempcnta\else
   {\advance\@tempcnta\@ne\ifnum\@tempcnta=\@tempcntb \else \def\@citea{--}\fi
    \advance\@tempcnta\m@ne\the\@tempcnta\@citea\the\@tempcntb}\fi\fi}
\catcode`@=12

\begin{document}

\title{\vskip-3cm{\baselineskip14pt
\centerline{\normalsize DESY~01-075\hfill ISSN 0418-9833}
\centerline{\normalsize hep-ph/0106135\hfill}
\centerline{\normalsize June 2001\hfill}}
\vskip1.5cm
Nonabelian $\alpha_s^3/(m_qr^2)$ heavy-quark-antiquark potential}
\author{{\sc Bernd A. Kniehl}, {\sc Alexander A. Penin}\thanks{Permanent
address: Institute for Nuclear Research, Russian Academy of Sciences,
60th October Anniversary Prospect 7a, Moscow 117312, Russia.},
{\sc Matthias Steinhauser}\\
{\normalsize II. Institut f\"ur Theoretische Physik, Universit\"at Hamburg,}\\
{\normalsize Luruper Chaussee 149, 22761 Hamburg, Germany}\\
\\
{\sc Vladimir A. Smirnov}\\
{\normalsize Nuclear Physics Institute, Moscow State University,}\\
{\normalsize 119899 Moscow, Russia}}

\date{}

\maketitle

\thispagestyle{empty}

\begin{abstract}
We calculate the leading mass-suppressed two-loop contribution, proportional
to $\alpha_s^3/(m_qr^2)$, to the potential of a heavy quark-antiquark system.
This contribution originates in the nonabelian nature of quantum
chromodynamics (QCD) and has no analogue in quantum electrodynamics (QED).
For this purpose, we elaborate a technique based on the implementation of the
threshold expansion within the effective theory of nonrelativistic QCD
(NRQCD).
We discuss phenomenological implications of our result for heavy-quarkonium
physics.
We also confirm a previous result for the two-loop ${\cal O}(\alpha_s^2)$
correction to the static potential.
\medskip

\end{abstract}

\newpage

Since the pioneering work by Susskind way back in 1977 \cite{Sus}, the static
potential between a heavy quark and its antiquark has attracted the attention
of many authors \cite{Fis,ADM,GupRad,TitYnd,Man,Pet,Sch}.
The structure of the heavy-quark-antiquark potential is crucial both for
understanding fundamental properties of QCD, such as confinement, and for
describing the rich phenomenology of heavy quarkonia.
While the potential, of course, cannot be found in a closed form, it can be
systematically computed as an expansion in the strong-coupling constant
$\alpha_s$ and the inverse of the heavy-quark mass $m_q$ or, equivalently, in
the heavy-quark velocity $v$.
The leading-order (LO) approximation is given by the simple generalization of
the Coulomb potential of QED.
The purely relativistic ${\cal O}(v^2)$ correction is given, up to the color
factor, by the standard Breit Hamiltonian.
The one-loop ${\cal O}(\alpha_s)$ correction has been known for a long time
\cite{Fis}, while the two-loop ${\cal O}(\alpha_s^2)$ correction has been
computed only recently \cite{Pet,Sch}.
The leading and subleading mass-suppressed one-loop ${\cal O}(v\alpha_s)$ and
${\cal O}(v^2\alpha_s)$ corrections have been found in
Refs.~\cite{GupRad,TitYnd,Man}.
In this Letter, we compute the two-loop leading mass-suppressed
${\cal O}(v\alpha_s^2)$ term.

Let us take a closer look at this specific class of corrections.
Owing to the gluon self-coupling, the QCD potential differs from its abelian
counterpart in two principal respects, namely by the running of the coupling
constant \cite{Sus} and the presence of generic corrections non-analytic in
$v^2$.
In general, terms non-analytic in $v^2$ can also appear in QED as a result of
iterations of operators that vanish on shell, in connection with the Coulomb
singularity, which brings in a factor of $1/v$.
One can avoid such terms by including off-shell operators in the Hamiltonian,
which is the standard practice in QED bound-state calculations.
The non-analytic terms that we consider here are a consequence of the
nonabelian structure of QCD, and they are not necessarily related to the
Coulomb singularity.
They first appear at one loop as the well-known {\it nonabelian}
$\alpha_s^2/(m_qr^2)$ potential \cite{GupRad}.
In this Letter, we take the next step and calculate this potential at two
loops.
Our main result is given by Eqs.~(\ref{2loopmom})--(\ref{2loopcoor}).
In order to derive it, we elaborate a technique based on the effective theory
of NRQCD \cite{CasLep} implemented with the threshold expansion offering the
possibility to systematically expand Feynman diagrams near threshold
\cite{BenSmi}.
We expect this approach to be very useful for solving a wide class of problems
of nonrelativistic dynamics.

The perturbative QCD potential is of special interest, since it is the basic
ingredient for the theory of heavy quarkonia.
In fact, the study of nonrelativistic heavy quark-antiquark systems
\cite{AppPol} and its applications to bottomonium \cite{NSVZ} and top-antitop
\cite{FadKho} physics rely entirely on first principles of QCD.
In principle, these systems allow for a perturbative treatment, the
nonperturbative effects \cite{VolLeu} being well under control and model
dependence being immaterial.
This makes heavy quark-antiquark systems to be an ideal laboratory to
determine the fundamental parameters of QCD, such as $\alpha_s$ and $m_q$.
Recently, essential progress has been made in the theoretical investigation of
the nonrelativistic heavy-quark threshold dynamics based on the
effective-theory approach \cite{CasLep}.
The analytical results for the main parameters of the nonrelativistic heavy
quark-antiquark system are now available through next-to-next-to-leading order
(NNLO) in $\alpha_s$ and $v$
\cite{PinYnd,CzaMel,HoaTeu,KPP,MelYel}, and they have been applied to
bottomonium \cite{KPP,MelYel,BenSin} and top-antitop \cite{gang}
phenomenology.
Some specific classes of next-to-next-to-next-to-leading order (N$^3$LO)
corrections have also been investigated \cite{KniPen1,BPSV,HMST};
for a brief review, see Ref.~\cite{Pen}.
These corrections have turned out to be so sizeable that it appears to be
indispensable to gain full control over this order, both in regard of
phenomenological applications and in order to understand the structure and the
peculiarities of the nonrelativistic expansion.
This Letter provides a major step in this direction.

In the following, we briefly outline the main features of our method, which is
based on the combination of the concepts of threshold expansion \cite{BenSmi}
and effective theory \cite{CasLep}.
The effective theory is designed to separate the scales and to perform the
expansion in $v$ at the level of the Lagrangian.
Let us recall that the dynamics of a nonrelativistic quark-antiquark pair
involves  four different scales, or {\it regions} \cite{BenSmi}:
(i) the hard scale (energy and three-momentum scale like $m_q$);
(ii) the soft scale (energy and three-momentum scale like $m_qv$);
(iii) the potential scale (energy scales like $m_qv^2$, while three-momentum
scales like $m_qv$); and
(iv) the ultrasoft scale (energy and three-momentum scale like $m_qv^2$).
The ultrasoft scale is only relevant for gluons.
NRQCD \cite{CasLep} is obtained by integrating out the hard modes.
Subsequently integrating out the soft modes and the potential gluons results
in the effective theory of potential NRQCD (pNRQCD) \cite{PinSot1}, which
contains potential quarks and ultrasoft gluons as active particles.
The dynamics of a nonrelativistic quark-antiquark pair in pNRQCD is governed
by the effective Schr\"odinger equation and by its multipole interactions with
the ultrasoft gluons.
The corrections from the modes integrated out
are contained in the higher-dimensional
operators of the nonrelativistic Hamiltonian, corresponding to an expansion in
$v$, and in the Wilson coefficients, which are expanded in $\alpha_s$.
Both NRQCD and pNRQCD have specific Feynman rules, which can be used for the
computation of a systematic perturbative expansion.
However, this problem is complicated because the expansion of the Lagrangian
corresponds to a particular subspace of the total phase space.
Thus, in a perturbative calculation within the effective theory, one has to
formally impose some restrictions on the allowed values of the virtual
momenta.
Explicitly separating the phase space introduces additional scales to the
problem, such as momentum cutoffs, and makes the approach much less
transparent.
A much more efficient and elegant method is based on the expansion by regions
\cite{BenSmi,Smi00}, which is a systematical method to expand Feynman diagrams
in any limit of momenta and masses.
It consists of the following steps:
(i) consider various regions of a loop momentum $k$ and expand, in every
region, the integrand in Taylor series with respect to the parameters that are
considered small there;
(ii) integrate the expanded integrand over the whole integration domain of the
loop momenta;
(iii) put to zero any scaleless integral.
In step (ii), dimensional regularization, with $d=4-2\epsilon$ space-time
dimensions, is used to handle the divergences.
In the case of the threshold expansion in $v$, one has to deal with the four
regions and their scaling rules enumerated above.

In principle, the threshold expansion has to be applied to the Feynman
diagrams of full QCD.
However, as we are only interested in the soft and potential contributions, it
is possible to apply step (i) to the diagrams constructed from the NRQCD
Feynman rules.
Equivalently, the Lagrangian of the effective theory can be employed for a
perturbative calculation without explicit restrictions on the virtual momenta
if dimensional regularization is used and the formal expressions derived from
the Feynman rules of the effective theory are understood in the sense of the
threshold expansion.
The last remark is crucial because, in general, the naive use of the
effective-theory Feynman rules and dimensional regularization leads to an
incorrect result.
Although this effect has not yet shown up in the QED bound-state calculations
performed along these lines so far \cite{PinSot2,KniPen3}, it is important for
our analysis.

As for the heavy-quark-antiquark potential, the contribution from the
hard  regions is analytic in $m_q$ and starts, for dimensional
reasons, at order $1/m_q^2$.
By definition, the effect of the ultrasoft modes should not be included in the
potential, so that we are left with the contributions of the soft and
potential regions.
In the effective-theory language, we study the reduction from NRQCD to pNRQCD
and compute the effect of the soft and potential modes being integrated out.
For this purpose, it is enough to use the Lagrangian of NRQCD.
Apart of the standard LO terms involving the gluon, ghost, quark, and
antiquark fields in the Lagrangian, we need to also include the
$1/m_q$-suppressed terms originating from the covariant-derivative operator
$\mbox{\boldmath$D$}^2/(2m_q)$ acting on the quark and antiquark fields.
Note that this operator also includes the quark kinetic-energy term
$\mbox{\boldmath$k$}^2/(2m_q)$.
It can be can be either treated as a perturbation if $k$ is soft or kept in
the nonrelativistic quark propagator
\begin{equation}
S(k)=\frac{1}{k_0-\mbox{\boldmath$k$}^2/(2m_q)+i\varepsilon}
\label{prop}
\end{equation}
if $k$ is potential.
This issue is discussed below in more detail.

By using the NRQCD Feynman rules, we immediately recover the well-known
one-loop result,
\begin{equation}
{\cal V}_{1/m_q}^{(1)}(\mbox{\boldmath$q$})
=\frac{\pi^2\alpha_s^2C_F}{m_q|\mbox{\boldmath$q$}|}
\left(\frac{C_F}{2}-C_A\right),
\label{1loopmom}
\end{equation}
where $C_F=(N^2-1)/(2N)$ and $C_A=N$ are the eigenvalues of the quadratic
Casimir operators of the fundamental and adjoint representations of the
$SU(N)$ color group, respectively.
In coordinate space, Eq.~(\ref{1loopmom}) takes the form
\begin{equation}
{\cal V}_{1/m_q}^{(1)}(\mbox{\boldmath$r$})
=\frac{\alpha_s^2C_F}{2m_q\mbox{\boldmath$r$}^2}
\left(\frac{C_F}{2}-C_A\right).
\label{1loopcoor}
\end{equation}
The abelian part enters the one-loop expression because we use the on-shell
potential.
In fact, we shall need it for the renormalization of our two-loop result.
In the standard QED analysis using the Coulomb gauge, it is removed by
introducing in turn the tree-level operator
\begin{equation}
{\cal V}_{\mathrm off}=\frac{\pi\alpha_sC_F}{m_q^2}\,
\left(\frac{\mbox{\boldmath$p^\prime$}^2-\mbox{\boldmath$p$}^2}
{\mbox{\boldmath$q$}^2}\right)^2,
\label{offshell}
\end{equation}
where {\boldmath$p$} and {\boldmath$p^\prime$} are the relative three-momenta
of the quark and antiquark, respectively, and
$\mbox{\boldmath$q$}=\mbox{\boldmath$p^\prime$}-\mbox{\boldmath$p$}$.
This is a purely {\it off-shell} operator, which vanishes for on-shell quarks,
when $\mbox{\boldmath$p$}^2=\mbox{\boldmath$p^\prime$}^2=2m_qE$, where $E$ is
the quark energy counted from the threshold.
By using the Coulomb equation of motion, it is straightforward to check that
the matrix elements of the abelian part of Eq.~(\ref{1loopmom}) and of
Eq.~(\ref{offshell}) between Coulomb states are the same.
The use of off-shell operators is advantageous in QED because it allows one to
reduce the number of loops by means of the Coulomb equation of motion, as may
be seen by comparing Eqs.~(\ref{1loopmom}) and (\ref{offshell}).
However, we use the on-shell formulation and the general covariant gauge,
which is more suitable for multiloop QCD calculations.

The calculation of the nonabelian part of Eq.~(\ref{1loopmom}) can be greatly
simplified by observing that the propagator of Eq.~(\ref{prop}) can be
expanded in $1/m_q$ not only in the soft region, but also in the potential
one, which is related to the contribution of its pole to the integral over
$k_0$.
This is possible because this part is free of Coulomb singularities.
In fact, performing the expansion, one recovers the familiar generalized
functions of $k_0$, $\delta^{(n)}(k_0)$.
By contrast, the abelian part suffers from the Coulomb singularity
corresponding to the iterations of the off-shell operator of
Eq.~(\ref{offshell}) and the Coulomb potential.
Note that the iterations of the operators included in the effective
Hamiltonian are taken into account by solving the Schr\"odinger equation
perturbatively around the Coulomb solution.
However, since we do not include the off-shell operator of
Eq.~(\ref{offshell}) in the effective Hamiltonian, we must include the result
of its iterations directly in the potential.
After expanding the quark propagator, this produces ill-defined products like
$1/[(k_0+i\varepsilon)^m(k_0-i\varepsilon)^n]$, which must be avoided by
resorting to the unexpanded version of Eq.~(\ref{prop}) in the potential
region.

The structure of the expansion remains intact at two loops, and our final
result reads
\begin{eqnarray}
{\cal V}_{1/m_q}^{(2)}(\mbox{\boldmath$q$})&=&
{\pi\alpha_s^3C_F\over m_q|\mbox{\boldmath$q$}|}
\left\{\left[\frac{(C_F-2C_A)\beta_0}{4}
\right.\right.\nonumber\\
&&{}-\left.\left.
\frac{4(2C_F+C_A)C_A}{3}\right]\ln\frac{\mu^2}{\mbox{\boldmath$q$}^2}+b_2
\right\},
\label{2loopmom}
\end{eqnarray}
where
\begin{eqnarray}
b_2&=&\left(\frac{65}{18}-\frac{8}{3}\ln2\right)C_FC_A-\frac{2}{9}C_FT_Fn_l
\nonumber\\
&&{}-\left(\frac{101}{36}+\frac{4}{3}\ln{2}\right)C_A^2
+\frac{49}{36}C_AT_Fn_l,
\label{b2}
\end{eqnarray}
$\beta_0=11C_A/3-4T_Fn_l/3$ is the one-loop coefficient of the QCD $\beta$
function, $T_F=1/2$ is the index of the fundamental representation, and $n_l$
is the number of light-quark flavors.
For $N=3$ and $n_l=3$, 4, and 5, we have $b_2\approx-20.836$, $-18.943$, and
$-17.049$, respectively.
In Eq.~(\ref{2loopmom}), we omitted the purely Abelian part, since the QED
potential to this order is known in the off-shell form.
The result evaluated in dimensional regularization may be found, for example,
in Ref.~\cite{PinSot2}.
In coordinate space, Eq.~(\ref{2loopmom}) becomes
\begin{eqnarray}
{\cal V}_{1/m_q}^{(2)}(\mbox{\boldmath$r$})&=&
\frac{\alpha_s^3C_F}{2\pi m_q\mbox{\boldmath$r$}^2}
\left\{\left[\frac{(C_F-2C_A)\beta_0}{4}
\right.\right.\nonumber\\
&&{}-\left.\left.
\frac{4(2C_F+C_A)C_A}{3}\right]
\ln\left(\tilde\mu^2\mbox{\boldmath$r$}^2\right)+b_2\right\},
\label{2loopcoor}
\end{eqnarray}
where $\tilde\mu=e^{\gamma_E}\mu$, with $\gamma_E$ being Euler's constant.
In our two-loop analysis, we again used the expanded form of Eq.~(\ref{prop})
in the calculation of the maximal nonabelian structures, $C_FC_A^2$ and
$C_FC_AT_Fn_l$, while it was necessary to keep Eq.~(\ref{prop}) unexpanded
for the treatment of the Coulomb singularities associated with the $C_F^2C_A$
and $C_F^2T_Fn_l$ terms.

We performed a number of nontrivial checks for our analysis.
(i) We worked in the general covariant gauge and verified that the gauge
parameter cancels in our final result.
(ii) The two-loop expression from which Eq.~(\ref{2loopmom}) is obtained
contains both ultraviolet (UV) and infrared (IR) divergences.
The UV ones were removed in Eq.~(\ref{2loopmom}) by the renormalization of
$\alpha_s$ in the one-loop result of Eq.~(\ref{1loopmom}), which we performed
in the modified minimal-subtraction ($\overline{\mathrm MS}$) scheme.
By the same token, the renormalization group logarithms proportional to
$\beta_0$ in Eq.~(\ref{2loopmom}) compensate the $\mu$ dependence of
Eq.~(\ref{1loopmom}).
On the other hand, the IR divergences are canceled by the UV ones of the
ultrasoft contribution \cite{KniPen1} leaving a finite, $\mu$-independent
result for the spectrum.
(iii) To test our program we also recalculated the two-loop correction to the
static heavy-quark-antiquark potential and found agreement with
Ref.~\cite{Sch}.
This is useful in its own right, since Ref.~\cite{Sch} disagrees with the
original result of Ref.~\cite{Pet}.

In Eq.~(\ref{2loopmom}), the IR divergences were subtracted according to the
$\overline{\mathrm MS}$ prescription.
For consistency, the same prescription must be used for the calculation of the
ultrasoft contribution.
We should mention that, in Ref.~\cite{KniPen1}, a slightly different
regularization/subtraction scheme was used.
There, only the divergent ultrasoft integrals were dimensionally regularized,
while the remaining convergent integrals over the potential region were solved
in four dimensions, despite overall factors containing poles in $\epsilon$.
This means that an additional matching term must be added to the result of
Ref.~\cite{KniPen1} so as to convert it into the conventional
$\overline{\mathrm MS}$ scheme used in the present paper.
In particular, this term contains logarithms that exactly cancel one half of
the IR logarithms in Eq.~(\ref{2loopmom}).
The remaining IR $\mu$ dependence is canceled by the ultrasoft logarithms,
involving the scale $E$ instead of $|\mbox{\boldmath$q$}|$, that result in
the logarithmically enhanced corrections to the bound-state parameters
\cite{BPSV}.

Our result is very relevant for heavy-quarkonium spectroscopy.
In fact, the complete N$^3$LO analysis of the spectrum includes three basic
ingredients:
(i) the leading retardation effect;
(ii) the multiple iterations of the low-order corrections to the potential;
and (iii) the N$^3$LO potential, which includes ${\cal O}(v^2\alpha_s)$,
${\cal O}(v\alpha_s^2)$, and ${\cal O}(\alpha_s^3)$ terms.
The retardation effect related to the radiation and absorption of ultrasoft
gluons, which results in the QCD analogue of the well-known Bethe logarithms,
was analyzed in Ref.~\cite{KniPen1}.
The corrections to the spectrum due to the multiple iterations of the
low-order corrections to the potential can be obtained using the
technique of Ref.~\cite{KPP}.
So far, only the one-loop ${\cal O}(v^2\alpha_s)$ part of the N$^3$LO
potential has been available \cite{GupRad,TitYnd,Man}.
Now that also the ${\cal O}(v\alpha_s^2)$ part is known, only the three-loop
${\cal O}(\alpha_s^3)$ correction to the static potential remains to be
computed.
Although the latter analysis is tedious from the technical point of view, it
is rather straightforward from the conceptual one.
We believe that the calculation of the two-loop ${\cal O}(v\alpha_s^2)$ term
performed here is probably most involved conceptually in the sense that the
full power of modern technology in effective field theory is required.

Our result is also very important in view of the recent analysis of
Ref.~\cite{HMST}.
There, it was conjectured that the perturbation theory for the normalization
of the $t\bar t$ production cross section at threshold can be essentially
improved by reordering the corrections and converting the standard
perturbative expansion in powers of $\alpha_s$ and $v$ around the Coulomb
solution to a logarithmic one, which sums up the infinite number of
logarithmically enhanced terms in each order by solving the nonrelativistic
renormalization group equation.
This conjecture is supported by an analysis involving the known part of the
NNLO anomalous dimensions, which is, however, incomplete.
The remaining part of the next-to-next-to-leading logarithmic (NNLL)
corrections can be obtained using an approach \cite{KniPen3} developed for the
calculation of the subleading logarithmic corrections in the spectrum of
positronium bound states.
The most nontrivial ingredient necessary to complete the calculation of the
NNLO anomalous dimensions is the two-loop result given in the present paper.
Note that the QED analysis \cite{KniPen3} implies that the missed terms are
likely to be important.
In fact, their abelian counterparts are responsible for the third-order
subleading logarithmic corrections to the positronium spectrum \cite{KniPen3}.
If the complete analysis supported the conjecture of Ref.~\cite{HMST}, this
would be crucial for the precision study of Higgs-boson-induced effects in
$t\bar t$ threshold production.

\vspace{1cm}
\noindent
{\bf Acknowledgements}
\smallskip

\noindent
The work of V.A.S. was supported in part by the Russian Foundation for Basic
Research through Project No.\ 01--02--16171.
This work was supported in part by the Deutsche Forschungsgemeinschaft through
Grant No.\ KN~365/1-1, by the Bundesministerium f\"ur Bildung und Forschung
through Grant No.\ 05~HT9GUA~3, and by the European Commission through the
Research Training Network {\it Quantum Chromodynamics and the Deep Structure
of Elementary Particles} under Contract No.\ ERBFMRX-CT98-0194.

\end{document}